\documentclass{wsmod}

\newcommand{\rr}{\mbox{\boldmath $r$}}

\newcommand{\rk}{\mbox{\boldmath $k$}}
\newcommand{\rkn}{\mbox{$k$}}

\newcommand{\rb}{\mbox{\boldmath $b$}}
\newcommand{\rp}{\mbox{\boldmath $p$}}

\newcommand{\be}{\begin{equation}}
\newcommand{\ee}{\end{equation}}
\newcommand{\beeq}{\begin{eqnarray}}
\newcommand{\eeeq}{\end{eqnarray}}

\begin{document}

\markboth{V. P. Gon\c{c}alves and M. V. T. Machado}
{Parton saturation approach in heavy quark production at high energies}

%%%%%%%%%%%%%%%%%%%%% Publisher's Area please ignore %%%%%%%%%%%%%%%
%
\catchline{}{}{}{}{}
%
%%%%%%%%%%%%%%%%%%%%%%%%%%%%%%%%%%%%%%%%%%%%%%%%%%%%%%%%%%%%%%%%%%%%

\title{Parton saturation approach in heavy quark production at high energies}

\author{V.P. Gon\c{c}alves}

\address{Instituto de F\'{\i}sica e Matem\'atica, Universidade Federal de
Pelotas\\
Caixa Postal 354, CEP 96010-090, Pelotas, RS, Brazil
\\
barros@ufpel.edu.br}

\author{M.V.T. Machado$^{a,b}$}

\address{$^a$ Universidade Estadual do Rio Grande do Sul - UERGS \\
 Unidade Bento Gon\c{c}alves, CEP 95700-000, Bento Gon\c{c}alves, RS, Brazil\\
$^b$ High Energy Physics Phenomenology Group, GFPAE  IF-UFRGS \\
Caixa Postal 15051, CEP 91501-970, Porto Alegre, RS, Brazil\\
magno-machado@uergs.edu.br, magnus@if.ufrgs.br
}

\maketitle

\pub{Received (Day Month Year)}{Revised (Day Month Year)}

\begin{abstract}

The  high parton density regime of the Quantum Chromodynamics
(QCD), where the physics of parton saturation is expected to be dominant, 
is briefly discussed. Some phenomenological aspects of
saturation are described, mainly focusing  on possible signatures
of the non-linear QCD dynamics in the heavy quark production in  electron-proton/nucleus
collisions. Implications of these effects in the heavy quark production in
ultraperipheral heavy-ion collisions are also presented.

\end{abstract}

\ccode{PACS Nos.: 12.38.-t; 13.85.-t; 24.85.+p}

\section{Introduction}

The understanding and analytic description of the  QCD at high
energies (small Bjorken $x$) has become  an increasingly active subject of
research, both from experimental and theoretical points of view (For a recent review, see e.g. \cite{iancuraju}).  These studies are mainly  motivated by the violation of the
unitarity (or Froissart-Martin bound \cite{froi}, which states that $\sigma_{tot}<\mathrm{C}\ln^2(s)$ at asymptotically large energies $s$) by the solutions of the linear
perturbative DGLAP \cite{dglap} and BFKL \cite{BFKL} evolution
equations. Since these evolution equations predict that the cross
section rises obeying a power law on energy, new dynamical effects associated with
the unitarity restoration are expected to stop its further growth
\cite{GLR}. This expectation can be easily understood:
while for large momentum transfer $\rk_{\perp}$, the BFKL  equation
predicts that the mechanism $g \rightarrow gg$ populates the
transverse space with a large number of small size gluons per unit
of rapidity (the transverse size of a gluon with momentum
$\rk_{\perp}$ is proportional to $1/\rk_{\perp}$), for small
$\rk_{\perp}$ the produced gluons overlap and fusion processes, $gg
\rightarrow g$, are equally important. Considering the latter
process, the rise of the gluon distribution below a typical scale
is slowed down, restoring the unitarity. That typical scale is energy
dependent and is called  saturation scale  $Q_{\mathrm{sat}}$.  The
saturation momentum sets the critical transverse size for the
unitarization of the cross sections. In other words, unitarity is
restored by including non-linear corrections in the evolution
equations\cite{GLR}. Such effects are small for $\rk_{\perp}^2 >
Q_{\mathrm{sat}}^2$ and very strong for $\rk_{\perp}^2 <
Q_{\mathrm{sat}}^2$, leading to the saturation of the scattering
amplitude.
 The magnitude of $Q_{\mathrm{sat}}$ is associated to the
behavior of the gluon distribution at high energies, and some
estimates has been obtained. In general, the predictions are
$Q_{\mathrm{sat}}\sim 1$ GeV at HERA/RHIC and $Q_{\mathrm{sat}}\sim 2-3$ GeV at LHC
\cite{GBW,vicslope}.  In particular, it has been observed
that the $ep$ deep inelastic scattering (DIS) data at low $x$ can
be successfully described with the help of the saturation model
\cite{GBW}, which incorporates the main characteristics of the
high density QCD approaches \cite{bal,kov,iancu}.

On the other hand,  deep inelastic scattering on nuclei gives us a
new possibility to reach high-density QCD phase without requiring
extremely low values of $x$. The nucleus in this process serves as
an amplifier for nonlinear phenomena.
In order to understand this expectation and estimate the kinematic
region where the high densities effects should be present, we can
analyze the behavior of the function $\kappa
(x,Q^2) \equiv \frac{3 \pi^2 \alpha_s  }{2 Q^2 } \frac{ xg_A(x,Q^2)}{\pi R^2_A%
} $, where $xg_A$ is the gluon distribution on the target $A$ of transverse size  $R_A \propto A^{1/3}$ probed by a virtual probe of virtuality $Q^2$. Such a function represents the probability of gluon-gluon interaction
inside the parton cascade, and also is  denoted the packing factor
of partons in a parton cascade \cite{GLR}. Considering that
the condition $\kappa = 1$ specifies the critical line, which
separates between the linear (low parton density) regime $\kappa
\ll 1$ and the high
density regime $\kappa \gg 1$, we can define the saturation momentum scale $%
Q_{\mathrm{sat}}$ given by
$
 Q_{\mathrm{sat}}^2 (x\,; A) = \frac{3 \pi^2 \alpha_s  }{2 } \frac{
xg_A(x,Q_{\mathrm{sat}}^2(x\,;A))}{\pi R^2_A}$,
below which the gluon density reaches its maximum value
(saturates). At any value of $x$ there is a value of $Q^2 =
Q_{\mathrm{sat}}^2(x)$ in which the gluonic density reaches a sufficiently high
value that the number of partons stops to rise. This scale depends
on the energy of the process [$xg \propto x^{- \lambda}$ ($\lambda
\approx 0.3$)] and on the atomic number of the colliding nuclei
[$R_A \propto A^{\frac{1}{3}} \rightarrow Q_s^2 \propto
A^{\frac{1}{3}}$], with the saturation scale for nuclear targets
larger than for nucleon ones.  This result
motivates more extensive studies of nuclear collisions and, in
particular, of electron-nucleus collisions at high energies, where nuclear medium effects are reduced in comparison with
$AA$ collisions.

In what follows we present a brief review of some signatures of
the high parton density regime, with special emphasis on the heavy quark
production in the saturation scenario. In the next section, we
present a brief review of the saturation approaches for deep
inelastic scattering process. In section \ref{geocharm} we discuss
the property of geometric scaling in the inclusive charm
production. Moreover, in Section \ref{photoheavy}, we consider the
heavy quark production in photonuclear process. In Section
\ref{peripheral} the possibility of using  ultraperipheral heavy
ion collisions as a photonuclear collider is analyzed and some
predictions for the saturation effects in the heavy quark
production are presented. Finally, in Section \ref{end} we
summarize our main conclusions.

\section{Overview on the Saturation Approaches}
\label{ea}

We start from the space-time picture of the eletron-proton/nuclei
processes \cite{dipole}. The deep inelastic scattering $ep(A)
\rightarrow e + X$ is characterized by a large electron energy
loss $\nu$ (in the target rest frame) and an invariant momentum
transfer $q^2 \equiv - Q^2$ between the incoming and outgoing
electron such that $x = Q^2/2m_N \nu$ is fixed ($m_N$ is the target mass). In terms of Fock
states we then view the $ep(A)$ scattering as follows: the
electron emits a photon ($|e\!> \rightarrow |e\gamma\!>$) with
$E_{\gamma} = \nu$ and $p_{t \, \gamma}^2 \approx Q^2$, after the
photon splits into a $q \overline{q}$ ($|e\gamma\!> \rightarrow |e
q\overline{q}\!>$) and typically travels a distance $l_c \approx
1/m_N x$, referred as the coherence length, before interacting in
the target. For small $x$, the photon converts to a quark
pair at a large distance before its scattering.
Consequently, the space-time picture of the DIS in the target rest
frame can be viewed as the decay of the virtual photon at high
energy  into a quark-antiquark pair (color dipole), which subsequently interacts with the target. In the small $x$ region, the color dipole crosses the
target with fixed transverse distance $\rr_{\perp}$ between the
quarks. The interaction $\gamma^*p(A)$ is further factorized and
is given by \cite{dipole},
\begin{eqnarray}
\sigma_{L,T}^{\gamma^*p(A)}(x,Q^2)= \sum_f \int dz \,d^2\rr_{\perp}
|\Psi_{L,T}^{(f)}(z,\rr_{\perp},Q^2)|^2
\,\sigma_{dip}^{p(A)}(x,\rr_{\perp}),
\end{eqnarray}
where $z$ is the longitudinal momentum fraction of the quark of flavor $f$. The
photon wave functions $\Psi_{L,T}$ are determined from light cone
perturbation theory  (For a review see, e. g., Ref. \cite{PREDAZZI}).

The dipole hadron (nucleus) cross section $\sigma_{dip}$  contains
all information about the target and the strong interaction
physics. Currently, the most complete QCD based effective theory
that describes the physics of hadronic interactions at very high
energies is the  Color Glass Condensate (CGC) \cite{iancu}. This represents the
small $x$ gluonic components in a hadronic wavefunction and  is
named so since: {\it Color} stands for the charge carried by the
gluons; {\it Glass} stands for a clear separation of time scales
between the fast and slow degrees of the wavefunction; {\it
Condensate} stands for the high density of gluons which can reach
values of order ${\cal{O}} (1/\alpha_s)$. The regime of  a CGC is
characterized by the limitation on the maximum phase-space parton
density that can be reached in the hadron/nuclear wavefunction
(parton saturation) and very high values of the QCD field strength
$F_{\mu \nu} \approx 1/\alpha_s$. The large values  of
the gluon distribution at saturation   suggest the use of
semi-classical methods, which allow to describe the small-$x$
gluons inside a fast moving nucleus by a classical color field \cite{mv}. In
the CGC  formalism \cite{iancu}, $\sigma_{dip}$ can be
computed in the eikonal approximation, resulting
\begin{eqnarray}
\sigma_{dip} (x,r_{\perp})=2 \int d^2 \rb_{\perp} \,\left[
1-\mathrm{S}\,(x,\rr_{\perp},\rb_{\perp})\right]\,\,,
\end{eqnarray}
where $\mathrm{S}$ is the $\mathrm{S}$-matrix element at fixed impact parameter $\rb_{\perp}$ which encodes all the
information about the hadronic scattering, and thus about the
non-linear and quantum effects in the hadron wave function. It can
be obtained by solving the functional  evolution equation in the
rapidity $y\equiv \ln (1/x)$ derived by Jalilian-Marian, Iancu,
McLerran, Weigert, Leonidov and Kovner (JIMWLK) \cite{iancu}. An equivalent
equation was developed by Balitsky \cite{bal}. These equations form a set of
coupled equations, and as such are very difficult to deal with
analitically. An approximated equation which allows to deal with
scattering at or near the unitarity limit was suggested by
Kovchegov\cite{kov}, and can be considered a mean field approximation for
the Balitsky-JIMWLK equations. While the Kovchegov equation is not
so complete it does have the advantage of being a precise
nonlinear equation for a function. Many interesting limits of the
Kovchegov equation has been understood by analytical methods, with
the main properties for the $\mathrm{S}$-matrix being: (a) for the
interaction of a small dipole ($\rr_{\perp} \ll 1/Q_{\mathrm{sat}}$),
$\mathrm{S}(\rr_{\perp}) \approx 1$, which characterizes that this system is
weakly interacting; (b) for a large dipole ($\rr_{\perp} \gg
1/Q_{\mathrm{sat}}$), the system is strongly absorbed which implies
$\mathrm{S}(\rr_{\perp}) \ll 1$.  This property is associate to the large
density of saturated gluons in the hadron wave function.

In
our analysis we will consider the  phenomenological
saturation model proposed in Ref. \cite{GBW} which encodes the
main properties of the  non-linear QCD approaches. In this model one has,
\begin{eqnarray}
\frac{\sigma_{dip}
(x,\rr_{\perp})}{\sigma_0}=1-\mathrm{S}\,(x,\rr_{\perp})\,;\,\,\,\,\,
\mathrm{S}=\exp\left[-\frac{Q_{\mathrm{sat}}^2(x)\,\rr_{\perp}^2}{4}\right],
\label{satgbw}
\end{eqnarray}
with $\sigma_{dip}/\sigma_0$ the scattering amplitude, averaged
over all impact parameters $\rb_{\perp}$, and
$Q_{\mathrm{sat}}^2\simeq\Lambda^2\,e^{\lambda\ln(x_0/x)}$. The
parameters of the model were constrained from the HERA small $x$
data, coming out  typical values of order 1-2 GeV$^2$ for the
momentum scale. We have that when
$Q_{\mathrm{sat}}^2(x)\,\rr_{\perp}^2\ll 1$, the model reduces to
color transparency, whereas as  one approaches the region
$Q_{\mathrm{sat}}^2(x)\,\rr_{\perp}^2 \approx 1$, the exponential
takes care of resumming many gluon exchanges, in a
Glauber-inspired way. Intuitively, this is what happens when the
proton starts to look dark.
Moreover, a  smooth transition to the
photoproduction limit is obtained with a modification of the
Bjorken variable as $x\rightarrow\tilde{x}= (Q^2 + 4\,m_f^2)/W_{\gamma p}^2$ and the large $x$ threshold corrections are accounted for by multiplying equation above by a factor $(1-x)^n$ [$n=5(7)$ for light (heavy) quarks].

One of the shortcomings of the saturation model is that the impact
parameters are averaged over. Thus there may be saturation at the
center of the proton, but the averaging gives much weight to the
edges of the proton, where saturation is not present. The implicit
assumption in the approach is that the proton is treated as being
homogeneous in the transverse plane. In such case, the impact
parameter profile is given by the Heaviside function,
$s(\rb_{\perp})=\Theta\,(b_0-b_{\perp})$, and is considered to be peaked at central
impact parameter, namely  at $\rb_{\perp}=0$. Actually, this procedure is
oversimplified and more realistic profiles can be considered. For
phenomenological purposes a Gaussian or a hard sphere assumption
are commonly taken into account. Recently, the impact parameter
dipole saturation model \cite{kowalski} was developed, recovering
the known Glauber-Mueller dipole cross section as well as the
DGLAP evolution has been included in the dipole cross section \cite{bartels}.
Using a simple parameterization for the $\mathrm{S}$-matrix element in
terms of the gluon distribution, those authors have obtained a
smooth matching onto the DGLAP evolution, at least in the leading
order DGLAP formalism, improving significantly the fit to the data
in the large $Q^2$ region.

Therefore, despite the saturation model to be very successful in
describing HERA data, its functional form is only an approximation
of the theoretical non-linear QCD approaches. Recently, a lot of work has been done to found analytical solutions from the saturation formalisms. Along these lines, currently intense
theoretical studies has been performed towards an understanding of
the BFKL approach in the border of the saturation region
\cite{IANCUGEO}. In particular, a parameterization
for the dipole cross section
 has been  implemented in Ref. \cite{iancu_munier}, where this quantity  was constructed to smoothly interpolate  between the  limiting behaviors analytically under control: the solution of the BFKL equation
for small dipole sizes, $\rr_{\perp}\ll 1/Q_{\mathrm{sat}}(x)$, and the
Levin-Tuchin law \cite{levin_tuchin} for larger ones, $\rr_{\perp}\gg
1/Q_{\mathrm{sat}}(x)$. The model has  been used in
phenomenological studies on  vector meson production
\cite{forshaw_vector},  diffractive processes \cite{forshaw_dif} and longitudinal structure function $F_L$\cite{goma_fl} at HERA,  as well as
 neutrino-nucleon total cross section \cite{magno_neu}.

\section{Heavy quarks in lepton-hadron collisions - Geometric Scaling}
\label{geocharm}

 An important feature of the available saturation
approaches is the prediction of the  geometric scaling. Namely,
the total $\gamma^* p$ cross section at large energies is not a
function of the two independent variables $x$ and $Q^2$, but is
rather a function of the single variable $\tau =
Q^2/Q_{\mathrm{sat}}^2$.  In Ref. \cite{IANCUGEO} the authors have demonstrated that
the geometric scaling predicted at low momenta $Q^2\leq
Q_{\mathrm{sat}}^2(x)$ is preserved by the BFKL evolution up to
relatively large virtualities, within the kinematical window
$Q_{\mathrm{sat}}^2 \leq Q^2 \ll
Q_{\mathrm{sat}}^4/\Lambda_{\mathrm{QCD}}^2$.  As demonstrated in
Ref. \cite{SGK}, the HERA data on the proton structure function
$F_2$ are consistent with scaling at $x \leq 0.01$ and $Q^2 \leq
400$ GeV$^2$. Similar behavior have been observed in exclusive
processes \cite{munier_wallon} and in the nuclear case
\cite{weigert_ea}.

In this section we analyze the geometric scaling in the inclusive
charm production \cite{vicmagprl}. From the  experimental point of view, the HERA
experiments have published data for the contribution of charmed
meson production to the structure function $F_2$. This allows one
to single out the charm contribution $F_2^c$ to the total
structure function and thus to investigate if the property of
geometric scaling is also present in this observable.  Before
presenting our results, lets perform a qualitative  analysis of
the inclusive charm production using the saturation model
\cite{GBW} in order to shed light on the dipole configurations
dominating the process in the relevant kinematical limits and show
how the geometric scaling comes out. A characteristic feature in
heavy quark production within the color dipole approach is that
the process is dominated by small size dipole configurations \cite{EPJCHQ}. The
overlap function weighting the dipole cross section is peaked at
$\rr_{\perp}\sim 1/m_c\simeq 0.1$ fm even for sufficiently low $Q^2$
values.  As a consequence, charm production is dominated by color
transparency and saturation effects are not important there, i.e.
$\sigma_{dip}\simeq \sigma_0\,Q_{\mathrm{sat}}^2(x) \rr^2_{\perp}/4$.

For the HERA kinematical region, we have $Q_{\mathrm{sat}}^2 \approx
1 $ GeV$^2$, which implies that the relation $Q_{\mathrm{sat}}^2<
Q^2 + \mu_c^2$ is ever satisfied, where $\mu_c^2\equiv 4m_c^2$. Consequently, we can define two
kinematical regimes depending of the relation between $Q^2$ and
$\mu_c^2$. For $Q^2 \gg \mu_c^2$ we have scaling with logarithmic
enhancement coming from aligned jet configurations, whereas for
$Q^2 \ll \mu_c^2$ only symmetric dipole configurations contribute.
Therefore, one obtains that the total cross section reads as,
\begin{eqnarray}
\sigma^{c\bar{c}}_{tot}   \sim \frac{\sigma_0\,Q_{\mathrm{sat}}^2(x)}{Q^2}
\left(1+ \ln \frac{Q^2}{\mu_c^2}  \right) \Theta (Q^2 - \mu_c^2) +
\frac{\sigma_0\,Q_{\mathrm{sat}}^2(x)}{\mu_c^2}\Theta
(\mu_c^2-Q^2)\,, \label{satexp}
\end{eqnarray}
where  the first term provides the behavior $1/\tau$ at large
$\tau$ whereas the second term leads to a smooth  transition down to
the  asymptotic ($\tau$-independent)  behavior at small $\tau$.

\begin{figure}[t]
\begin{center}
\begin{tabular}{cc}
\epsfig{file=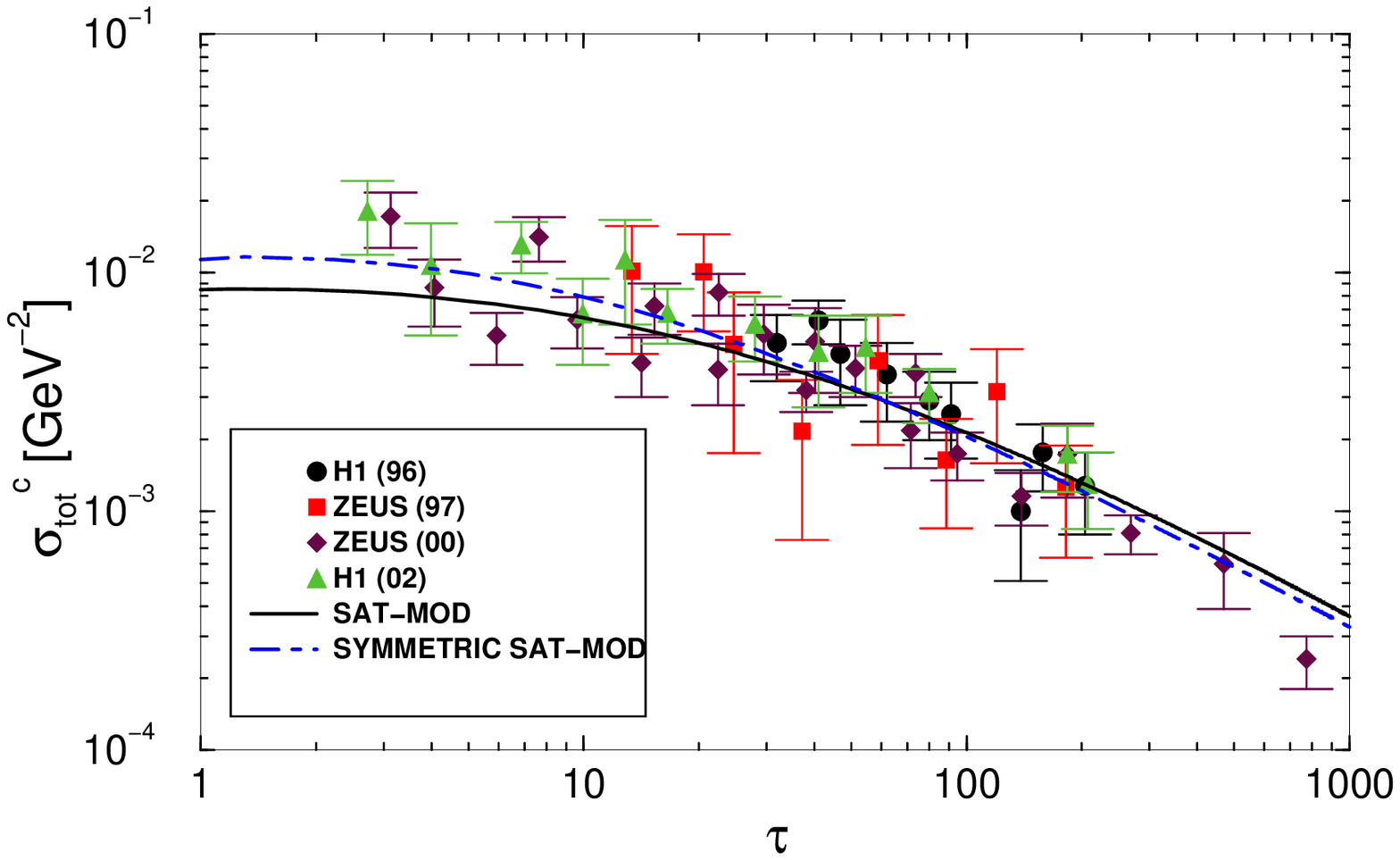,width=60mm} & \epsfig{file=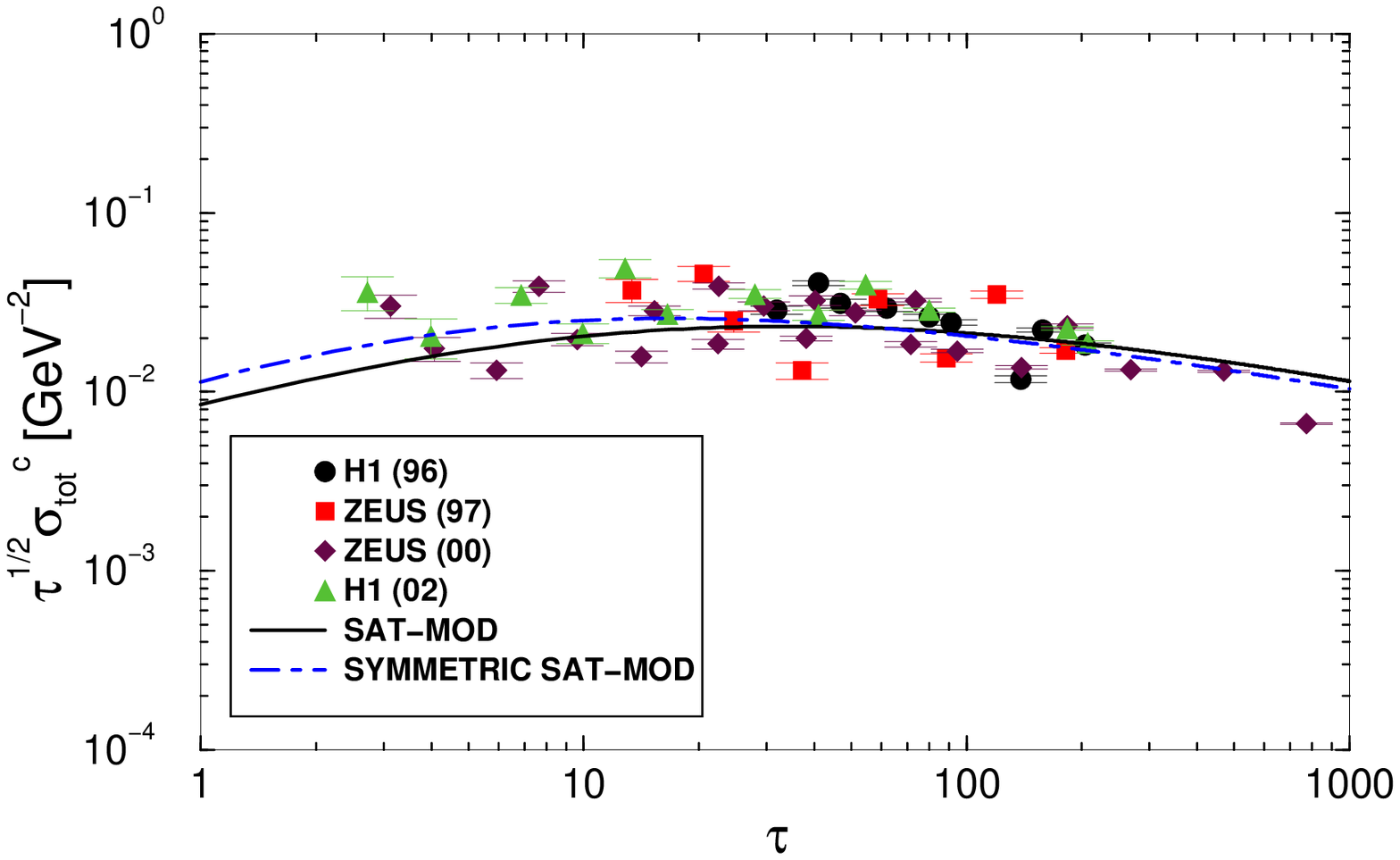,width=60mm}
\end{tabular}
\end{center}
\caption{Experimental data on inclusive charm
production plotted versus the scaling variable. The curves are the saturation model (solid line) and symmetric saturation model (dashed lines).}
%\end{tabular}
\label{fig:1}
\end{figure}

An analytical expression for  the $\tau$ dependence of the
inclusive charm production  can also be obtained in a less model
dependent way.  For this purpose we will make use of the symmetric
saturation model \cite{munier_sim}, where the energy evolution of the
proton leads to the parton multiplication and the transverse
momentum scale $Q_{\mathrm{sat}}(x) $ appears. The main assumption
is that the evolved proton can be described by a collection of
independent dipoles at the time of the interaction whose sizes are
distributed around $1/Q_{\mathrm{sat}}$. The rate of growth of the
parton densities is assumed to be $Q_{\mathrm{sat}}^2(x)/\Lambda^2$
and the symmetry between low and high virtualities in $\gamma p$
interactions comes from the symmetry in the dipole-dipole cross
section. In the HERA kinematical regime, i.e.
$Q_{\mathrm{sat}}^2<\mu_c^2$, the inclusive charm production is
given by,
\begin{eqnarray}
\sigma^{c\bar{c}}_{tot}\,(\tau)= \frac{N_{c\bar{c}}}{\Lambda^2\,\nu_{>}}\,\left\{ 1- \exp \left[ -\frac{\nu_{>}}{\tau + \tau_c}\left( 1+ \log (\tau + \tau_c\right)     \right]\right\}\,,
\end{eqnarray}
where $\tau_c= \mu_c^2/Q_{\mathrm{sat}}^2(x)$ and the parameters $N$, $\Lambda^2$ and $\nu_{(>,<)}$ are
taken from the data fit in Ref. \cite{munier_sim}. Here, we make the
simplified assumption that the coupling with the dipole is flavor
blind, in such way that $N_{c\bar{c}} = (2/5) N$, with $N$ being
the global  normalization describing $F_2$ data. The factor $2/5$
corresponds  to the charge fraction $e_c^2/(\sum e_{u,\,d,\,s}^2 +
e_c^2)$. Once the  parameters are fitted to proton structure
function data, our prediction for the $\tau$ dependence in the
inclusive charm production is parameter free.

In Fig.  \ref{fig:1}  we show  the experimental data on the
total cross section for the inclusive charm production plotted
versus scaling variable $\tau$, with
$Q_{\mathrm{sat}}$  from the saturation model.  We see the data
exhibit geometric scaling for the whole $Q^2$ range, verifying a
transition in the behavior on $\tau$  of the cross section from a
smooth dependence at small $\tau$ and an approximated $1/\tau$
behavior at large $\tau$. The transition point is placed at
$\mu_c^2=4m_c^2$, which takes values of order 10 GeV$^2$ for a
charm mass $m_c=1.5$ GeV. This turns out in $\tau \simeq 10$ since
at HERA $Q_{\mathrm{sat}}^2\simeq 1$ GeV$^2$. The asymptotic
$1/\tau$ dependence reflects the fact the charm production cross
section scales as $Q_{\mathrm{sat}}^2/Q^2$ modulo a logarithmic
correction $\sim \ln (Q^2/\mu_c^2)$, with energy dependence driven
by the saturation scale. The mild dependence at $\tau\leq \mu_c^2$
corresponds to the fact the cross section scales as
$Q_{\mathrm{sat}}^2/\mu_c^2$ towards  the photoproduction limit, but
with the same energy behavior given by the saturation scale. In
Fig. \ref{fig:1}, we also found a symmetry between the regions
of large and small $\tau$ for the function
$\sqrt{\tau}\,\sigma^{c\bar{c}}_{tot}$ with respect the
transformation $\tau \leftrightarrow 1/\tau$ in the whole region
of $\tau$. The features present in the inclusive charm production
data can be well reproduced in the phenomenological saturation
model, corresponding to the solid curve in Fig. \ref{fig:1}.
The symmetric saturation model also provides similar results, as
shown in the dot-dashed lines. Disregarding the Glauber-like
resummation in  this model, the  expression gets simplified to
$\sigma^{c\bar{c}}_{tot}\propto \frac{1}{\tau+\tau_c}\,[1+\log(\tau +
\tau_c)]$, and the  symmetric pattern is easily verified.

In the HERA kinematic domain the saturation momentum
$Q_{\mathrm{sat}}^2(x)$ stays  below the hard scale
$\mu_c^2=4m_c^2$, implying that charm production probes mostly the
color transparency regime   and saturation corrections are not
very important. However,  as the saturation scale rises with the
energy and the atomic number, we expect that at larger energies
and nuclear collisions a new  kinematic regime where $
Q_{\mathrm{sat}}^2(x) \ge  \mu_c^2$ will be probed. Recently, the
production of open charm in heavy ion collisions in the CGC
framework has been considered in Ref. \cite{kharzeev_tuchin}. The
main prediction is the approximate scaling of the cross section
with the number of participants ($N_{\mathrm{part}}$) in the
forward rapidity region, where the saturation scale exceeds the
charm quark mass. This result is in contrast with usual
expectation for a hard process of scaling with the number of
collisions ($N_{\mathrm{coll}}$). These results provide a strong
motivation for further investigations (See, e.g. Refs.
\cite{raju_gelis,tuchin_pa,raju_gelis2}).

\section{Heavy quark production in photonuclear process}
\label{photoheavy}

In this section, we report our investigations on the high energy
heavy quark photoproduction on nuclei targets using the saturation
hypothesis \cite{EPJCHQ}. In particular, we study them considering
the approach proposed in Ref. \cite{armesto}, which extends the
saturation model for scattering on nuclei and gives a reasonable
parameter-free description of the experimental data on nuclear
structure function. In this model the dipole-nucleus cross section
is given by \cite{armesto},
\begin{eqnarray}
\sigma_{dip}^{\mathrm{nucleus}} (x, \,\rr_{\perp}^2, A)  = 2 \int d^2\rb_{\perp} \,
\left\{\, 1- \exp \left[-\frac{1}{2}\,A\,T_A(\rb_{\perp})\,\sigma_{dip}^{\mathrm{nucleon}} (x, \,\rr_{\perp}^2)  \right] \, \right\}\,,
\label{sigmanuc}
\end{eqnarray}
where $\rb_{\perp}$ is the impact parameter of the center of the dipole
relative to the center of the nucleus and the integrand gives the
total dipole-nucleus cross section for a  fixed impact parameter.
The nuclear profile function is labeled by $T_A(\rb_{\perp})$. The above
equation sums up all the multiple elastic rescattering diagrams of
the $q \overline{q}$ pair and is justified for large coherence
length, where the transverse separation $\rr_{\perp}$ of partons in the
multiparton Fock state of the photon becomes as good a conserved
quantity as the angular momentum, {\it i. e.} the size of the pair
$\rr_{\perp}$ becomes eigenvalue of the scattering matrix. Here, the dipole
cross section for the nucleon target (proton),
$\sigma_{dip}^{\mathrm{nucleon}} (x, \,\rr_{\perp}^2)$, is given by the
saturation model [Eq. (\ref{satgbw})]. The photoproduction cross section reads as,
\begin{eqnarray}
\sigma_{tot}(\gamma A\rightarrow Q\overline{Q}X)=\int dz \,d^2\rr_{\perp}
|\Psi_{T}^{Q\overline{Q}}(z,\rr_{\perp},Q^2=0)|^2
\,\sigma_{dip}^{\mathrm{nucleus}}(x,\rr_{\perp})\,,
\end{eqnarray}
where the dipole-nucleus cross section is given by Eq. (\ref{sigmanuc}) and only the transverse wavefunction for the heavy quark-antiquark pair $Q\overline{Q}$ contributes at $Q^2\rightarrow 0$.

\begin{figure}[t]
\centerline{\psfig{file=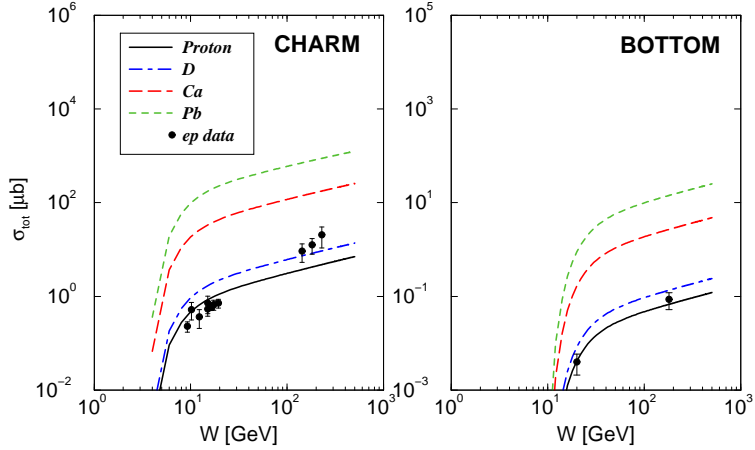,width=100mm}}
 \caption{\it The total nuclear cross section for charm and bottom photoproduction as
 a function of energy $W_{\gamma A}$ for distinct nuclei.}
 \label{fig1}
\end{figure}

In Fig. \ref{fig1} are shown the results for the charm and bottom
photoproduction cross section as a function of energy for
different nuclei, including the proton case. The results present
mild growth on $W_{\gamma A}$ at high energies stemming from the saturation
model, whereas the low energy region is consistently described
through the large-$x$ threshold factor \cite{EPJCHQ}. For the proton, the experimental data from HERA  and fixed target collisions  are also included for  comparison. The result for charm underestimates data by a factor 2 at $W_{\gamma p} \simeq 200$  GeV, whereas is consistent with the measurements of bottom cross section. Concerning charm production, the measured cross sections present a well known steeper behavior on energy even at electroproduction, suggesting that further resummations in the original saturation model are needed in order to produce the  larger growth on energy  appearing in the charm measurements. We believe that the better  result for bottom happens to be  a mismatch between a large uncertainty in the experimental measurement and the lower bottom mass $m_b=4.5$ GeV  considered here.   For the nuclear case,  we predict that their absolute values are rather large, reaching $\approx 2 \cdot 10^3$  and $\approx 40 \, \mu b$
for charm and bottom for lead at $W_{\gamma A} = 10^3$ GeV.

The simple intuitive dipole approach considered above has a deep connection with a more general theoretical formalism. Namely, it is equivalent, at leading logarithmic approximation, to the high energy  $k_{\perp}$-factorization (semihard) approach  \cite{smallx}. In this framework, the
relevant QCD diagrams are considered with the virtualities and
polarizations of the initial partons, carrying information on
their transverse momenta. The scattering processes are described
through the convolution of off-shell matrix elements with the
unintegrated parton distribution, ${\cal F}(x,\rk_{\perp})$. The characteristic feature is the LO cross section in this approach resumming most of the NLO and even NNLO diagrams contributing to the process in the collinear formalism. As heavy quark photoproduction is concerned, considering only the direct component of the photon, the cross section reads as \cite{smallx},
\begin{eqnarray}
\sigma_{tot}(\gamma\, A \rightarrow Q\overline{Q}X) & = & 
  \frac{\alpha_{em}\,e_Q^2}{\pi}\, \int\, dz\,\,d^2 \rp_{1\perp} \, d^2\rk_{\perp} \, \frac{\alpha_s(\mu^2)\,{\cal F}(x,\rk_{\perp}^2)}{\rk_{\perp}^2} \nonumber \\
& \times & \!\left\{[z^2+ (1-z)^2]\, \left( \frac{\rp_{1\perp}}{D_1} + \frac{(\rk_{\perp}-\rp_{1\perp})}{D_2} \right)^2 +   m_Q^2 \,\left(\frac{1}{D_1} + \frac{1}{D_2}  \right)^2  \right\}\,,\nonumber
 \label{sigmakt}
\end{eqnarray}
where  $D_1 \equiv \rp_{1\perp}^2 + m_Q^2$ and $D_2 \equiv
(\rk_{\perp}-\rp _{1\perp})^2 + m_Q^2$.
The transverse momenta of the heavy quark (antiquark) are denoted
by $\rp_{1\perp}$ and $\rp_{2\perp}= (\rk_{\perp}-\rp _{1\perp})$,
respectively. The heavy quark longitudinal momentum fraction is
labeled by $z$. For the scale $\mu$ in the strong coupling
constant we  use the prescription  $\mu^2=\rk_{\perp}^2 +
m_{Q}^2$. The unintegrated gluon distribution ${\cal F}(x,\rk_{\perp}^2)$  is directly related to the Fourier transform of the  dipole-nucleon (nucleus)
total cross section, as follows 
\begin{eqnarray}
\frac{{\cal F}(x,\rk_{\perp}^2)}{\rk_{\perp}^2}=
\frac{3}{4\pi\alpha_s} \int d^2\rb_{\perp}\int \frac{d^2 \rr_{\perp}}{(2\pi)^2} \, e^{\,i\,\rk_{\perp} \cdot \rr_{\perp}} \left[\sigma_{dip}
(x,\rr_{\perp} \rightarrow \infty,\rb_{\perp}) - \sigma_{dip} (x,\rr_{\perp},\rb_{\perp})\right]\,.\nonumber \label{ufgdef}
\end{eqnarray} 

For the saturation model this quantity takes a simple analytical form, which reads for the proton and nucleus case as \cite{among},
\begin{eqnarray}
{\cal{F}}^{\mathrm{sat}}_{\mathrm{proton}}(x,\rk_{\perp})& = & \frac{3\,\sigma_0}{4\pi^2\alpha_s}\,\left(\frac{\rk_{\perp}^2}{Q_{\mathrm{sat}}^2(x)}\right)\,\exp \left(-\frac{\rk_{\perp}^2}{Q_{\mathrm{sat}}^2(x)}\right) \,,\\
{\cal{F}}^{\mathrm{sat}}_{\mathrm{nucleus}}(x,\rk_{\perp},\rb_{\perp})& = & \frac{3}{2\pi^2\alpha_s}\,\left(\frac{\rk_{\perp}^2}{Q_{\mathrm{s}A}^2(x,\,b_{\perp})}\right)\,\exp \left(-\frac{\rk_{\perp}^2}{Q_{\mathrm{s}A}^2(x,\,b_{\perp})}\right)\,,
\end{eqnarray}
where the expression for nuclei is obtained under the assumption of dominance of color transparency (small dipole configurations) in the dipole cross section, which is correct for the particular case of heavy quark production. The nuclear saturation scale is given by $Q_{\mathrm{s}A}^2(x,b_{\perp})=\frac{1}{2}A\,T_A(b_{\perp})\,\sigma_0\,Q_{\mathrm{sat}}^2(x)$. We call attention to the scaling pattern on the variable $\tau=\rk_{\perp}^2/Q_{\mathrm{s}A}^2$, which implies scaling on $\tau$ also in the nuclear heavy quark production. In our further analysis, they will be  compared with the simple ansatz  ${\cal
F}_{\mathrm{nuc}} = \frac{\partial\,
xG_{A}(x,\,\rk_{\perp}^2)}{\partial \ln \rk_{\perp}^2}$, with $xG_{A}(x,Q^2)$ being  the nuclear gluon distribution (see
\cite{EPJCHQ,among} for details  in the  numerical calculations).

On the other hand,  in the collinear  approach the cross section is given by
a convolution between the partonic cross section for the
subprocess $\gamma g \rightarrow Q \overline{Q}$ and the
integrated gluon distribution for the nucleus $xG_A(x,Q^2)$. In
our studies in Ref. \cite{among}, one considers the  EKS \cite{EKS} and AG \cite{ayavic}
parameterizations for this distribution. The EKS parameterization
was obtained from a global fit of the nuclear experimental data
using the DGLAP evolution equations, which is a linear evolution
equation which does not consider dynamical saturation (high
density) effects. In Ref. \cite{ayavic} a procedure to include
these effects  in the nuclear gluon distribution was proposed,
resulting in a paramerization for this distribution (AG
parameterization), which also includes those present in the EKS
parameterization. The AG parameterization predicts a stronger
reduction of the growth of the gluon distribution at small values
of $x$ than the EKS one.

When the nuclear photoproduction cross section of heavy quarks is
computed considering these different approaches, we have obtained
that  the $k_{\perp}$-factorization using EKS unintegrated gluon
pdf (semihard approach) gives similar results to the collinear
approach where nuclear effects (EKS parameterization)
and high density corrections (AG parameterization)
are taken into account. In particular, we have that the
predictions using the AG parameterization in the collinear
approach are similar to the semihard one, which does not consider
high density effects. This demonstrate that in this process we
cannot distinguish if the modification in the behavior of the
cross section is associated to high density effects in the
collinear approach or a generalization of the factorization
without high density effects in the unintegrated gluon
distribution. On the other hand, if these effects are present and
the factorization of the cross section is given by the
$k_{\perp}$-factorization, as is the case for the predictions from
the saturation model, we have that the difference between the
cross sections is large, which should allow to discriminate
between the theoretical approaches.  Therefore, the nuclear cross
section would provide a strong test concerning the robustness of
the saturation approach in describing the observables. The
situation is less clear comparing the semihard approach and the
collinear one. One possible interpretation for this result is that
the expected enhancement in the semihard approach, associated to
the resummation of the $(\alpha_s \, \ln \frac{\sqrt{s}}{m_Q})^n$
in the coefficient function, is not sizeable for
inclusive quantities  in  the kinematic region of the future
colliders. Probably, a more promising quantity to clarify this
issue would be the transverse momentum $\rp_{\perp}$ distribution.
In this case, the semihard approach seems to be in better
agreement with experimental data in the $pp$ collisions than the
collinear approach \cite{smallx}.

\section{Heavy Quarks in Ultraperipheral Heavy Ion Collisions}
\label{peripheral}

The studies of saturation effects in nuclear
processes shown that future electron-nucleus colliders at HERA and
RHIC, probably could determine whether parton distributions
saturate and constrain the behavior of the nuclear gluon
distribution in the full kinematical range. However, until these
colliders become reality we need to consider alternative searches
in the current and/or scheduled accelerators which allow us to
constrain the QCD dynamics. Recently, we have analyzed the
possibility of using ultraperipheral heavy ion collisions (UPC's) as a
photonuclear collider. In
particular, we have studied the heavy quark production \cite{EPJCPER3} assuming distinct
approaches for the QCD evolution. 

In heavy ion  collisions the large number of photons coming from
one of the colliding nuclei  will  allow to study photoproduction,
with energies $W_{\gamma A}$ reaching to almost 1 TeV for the LHC. The
photonuclear cross sections are given by the convolution between
the photon flux from one of the nuclei and the cross section for
the scattering photon-nuclei, with the photon flux
$\frac{dN\,(\omega)}{d\omega}$ given by the Weizsacker-Williams
method \cite{BaurPR}.  The final  expression for the
production of heavy quarks in ultraperipheral heavy ion collisions
is  given by,
\begin{eqnarray}
\sigma_{AA \rightarrow
Q\overline{Q}X}\,\left(\sqrt{S_{\mathrm{NN}}}\right) = \int
\limits_{\omega_{min}}^{\infty} d\omega \,
\frac{dN\,(\omega)}{d\omega}\, \sigma_{\gamma A \rightarrow Q\overline{Q}X}
\left(W_{\gamma A}^2=2\,\omega\sqrt{S_{\mathrm{NN}}}\right)\,
\label{sigAA}
\end{eqnarray}
where $\omega$ is the c.m.s. photon energy, $\omega_{min}=M_{Q\overline{Q}}^2/4\gamma_L m_p$ and
$\sqrt{S_{\mathrm{NN}}}$ is  the c.m.s energy of the
nucleus-nucleus system. The Lorentz factor for LHC is
$\gamma_L=2930$, giving the maximum c.m.s. $\gamma N$ energy
$W_{\gamma A} \approx 950$ GeV.
 The requirement that  photoproduction
is not accompanied by hadronic interaction (ultraperipheral
collision) can be done by restricting the impact parameter $b$  to
be larger than twice the nuclear radius, $R_A=1.2 \,A^{1/3}$ fm. An analytic approximation for $AA$ collisions can be obtained
using as integration limit $b>2\,R_A$, producing
\begin{eqnarray}
\frac{dN\,(\omega)}{d\omega}= \frac{2\,Z^2\alpha_{em}}{\pi\,\omega}\, \left[\bar{\eta}\,K_0\,(\bar{\eta})\, K_1\,(\bar{\eta})+ \frac{\bar{\eta}^2}{2}\,\left(K_1^2\,(\bar{\eta})-  K_0^2\,(\bar{\eta}) \right) \right] \,,
\label{fluxint}
\end{eqnarray}
where $\bar{\eta}=2\omega\,R_A/\gamma_L$ and $K_{0,1}(x)$ are the modified Bessel functions.
The typical values of  $x$ which
will be probed in ultraperipheral heavy ion collisions, are given
by $x = (M_{Q\overline{Q}}/2p) e^{-y}$, where $M_{Q\overline{Q}}$
is the invariant mass of the photon-gluon system and $y$ the
center of momentum rapidity. For Pb + Pb collisions at LHC
energies the nucleon momentum is equal to $p=2750$ GeV; hence $x =
(M_{Q\overline{Q}}/5500 \, {\rm GeV}) e^{-y}$. Therefore, the
region of small mass and large rapidities probes directly the high
energy (small $x$) behavior of the QCD dynamics present in the
$\gamma \, A$ cross section. This demonstrates that
ultraperipheral heavy ion collisions at LHC represents a very good
tool to constrain the high energy regime of the QCD dynamics.

In what follows we summarize our  analysis on photonuclear
production of heavy quarks at UPC's \cite{EPJCPER3}. In order to  do this, one
considers the available high energy approaches. Namely, the usual
collinear approach, the semihard formalism and  the
phenomenological saturation reviewed in the previous section.
As an additional analysis, we also consider the Color Glass
Condensate (CGC) formalism. This puts into  perspective the most
recent representative high energy approaches and allow us to find
out the observables and/or distributions which could disentangle
them at the planned colliders.

In Refs. \cite{gelis} the heavy quark production  in UPC's has
been analyzed in the CGC formalism.  In Ref. \cite{EPJCPER3}, we
have improved that analysis using a realistic photon flux and a
color field correlator including quantum radiation effects. The
differential cross section on rapidity reads as \cite{EPJCPER3},
\begin{eqnarray}
\frac{d\sigma_{AA\rightarrow Q\overline{Q}X}}{dY}  = \omega \,
\frac{dN(\omega)}{d\omega}\,\frac{\alpha_{em}e_Q^2}{2\,\pi}
\int\limits_{0}^{+\infty} d\rk_{\perp}^2\, R_A^2\,
\widetilde{C}\,(\rk_{\perp}) \, \left\{ 1+
\frac{4(\rk_{\perp}^2-m_Q^2)}{\rkn_{\perp}\bar{\mu}_Q^2}\,{\rm
arcth}\,\frac{\rkn_{\perp}}{\bar{\mu}_Q^2}
\right\}\,, \nonumber
\label{dsdy_phen}
\end{eqnarray}
where we define the  rapidity $Y\equiv \ln(1/x)=
\ln(2\,\omega\,\gamma_L/4m_Q^2)$. The quark charge is labeled as $e_Q$ and one uses the notation $\bar{\mu}_Q^2\equiv \sqrt{\rk_{\perp}^2+4m_Q^2}$. In Ref.
\cite{EPJCPER3}, we obtained the following analytical expression for the color field
correlator, considering that it  is directly related to the Fourier transform of the  dipole-nucleus
total cross section,
\begin{eqnarray}
\widetilde{C}\,(x,\rk_{\perp})= \left(\frac{4\pi}{Q_{\mathrm{s}A}^2(x)}\right)\, \exp \left( -\frac{\rk_{\perp}^2}{Q_{\mathrm{s}A}^2(x)} \right)\,,
\label{csat}
\end{eqnarray}
where we have assumed $Q_{\mathrm{s}A}^2 (x) = A^{1/3}\,Q_{\mathrm{sat}}^2
(x)$.  We believe that this input is more suitable for realistic
computations because it includes quantum evolution in the
formalism and reproduces most part of the phenomenological
features of the saturation model for the nucleus case. It is worth mentioning the direct relation between the color field correlator and the unintegrated gluon distribution, given by ${\cal{F}}(x,\rk_{\perp}) = (3 R_A^2 / 8 \pi^2 \alpha_s)\,\rk_{\perp}^2 \,C\, (x,\rk_{\perp})$.

\begin{figure}[t]
\begin{center}
\resizebox{!}{5cm}{\includegraphics{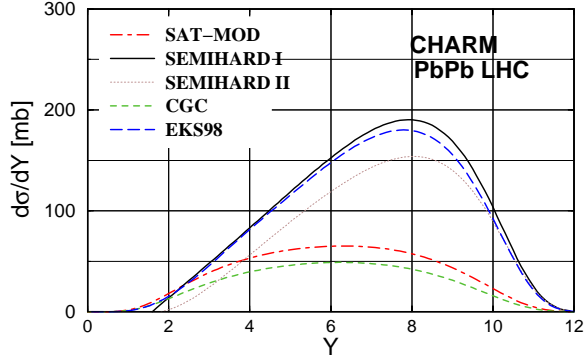}}
\end{center}
\caption{Rapidity distribution for charm production in
ultraperipheral heavy ion collisions.}\label{rapcharm}
\end{figure}

In Fig. \ref{rapcharm} is shown the charm rapidity distribution  for
the distinct high energy approaches considered.  The collinear
result is denoted by the long-dashed curves, where we have
employed the EKS98 parameterization for the collinear nuclear
gluon function (See also Ref. \cite{vicbert}). The solid and dotted lines label the semihard
($k_{\perp}$-factorization) results, where one has used  the
unintegrated gluon distribution  ${\cal F}_{\mathrm{nuc}}\,(x,\,k_{\perp}^2;\,A)$ discussed in the previous section. Two possibilities for the nucleon gluon distribution
were considered: (I) GRV94(LO) - solid line - and (II) GRV98(LO) -
dotted line. The saturation model results are denoted by the
dot-dashed line.  The CGC prediction is denoted by the
dashed line. We have that the predictions for the collinear
approach and the semihard formalism are similar  and give somewhat
larger values than the saturation and CGC results. One possible
interpretation for the similarity between the predictions of the
semihard approach and the collinear one is that the expected
enhancement in the $k_{\perp}$-factorization formalism
is not sizeable for inclusive quantities  in  the kinematic region
of the future colliders \cite{smallx}. Our phenomenological ansatz within the
CGC formalism gives similar results as the saturation model, but
should be noticed that the physical assumptions in those models
are distinct. While  the saturation model considers multiple
scattering on single nucleons, our expression for the
dipole-nucleus cross section in the CGC formalism assumes
scattering on a black area filled by partons coming from many
nucleons.

Let us    present the numerical calculation of their total cross
section at UPC's. We focus mostly on LHC domain where small values
of $x$ would be probed.  The results are presented in Table
\ref{tabhq}. The collinear approach gives a larger rate, followed
by the semihard approach. The saturation model and CGC formalisms
give similar results, including a closer ratio for charm to bottom
production. Concerning the CGC approach, our phenomenological
educated guess  for the color field correlator seems to produce
quite reliable estimates. Therefore, the photonuclear production
of heavy quarks allow us to constraint already in the current
nuclear  accelerators the QCD dynamics since the main features
from photon-nuclei collisions hold in the UPC reactions.
 Our
results shown that an experimental analysis of this process can be
useful to constrain the QCD dynamics at high energies.

\section{Summary}
\label{end}

 The perturbative QCD has furnished a remarkably
successful framework to interpret a wide range of high energy
lepton-lepton, lepton-hadron and hadron-hadron processes. Through
global analysis of these processes, detailed  information on the
parton structure of hadrons, especially the nucleon, has been
obtained. The existing global analysis have been performed using
the standard DGLAP evolution equations. However, in  the small $x$
region the DGLAP evolution equations are expected to breakdown,
since new dynamical effects associated to the high parton density
must occur in this kinematical region.

\begin{table}[t]
%\begin{center}
\tbl{\it The photonuclear heavy quark  total cross sections for UPC's  at LHC.}
{\begin{tabular} {||c|c|c|c|c||}
\hline
\hline
$Q\overline{Q}$   & {\bf Collinear} & {\bf SAT-MOD} & {\bf SEMIHARD I (II) } &  {\bf CGC} \\
\hline
 $c\bar{c}$ & 2056 mb & 862 mb &  2079 (1679.3) mb & 633 mb \\
\hline
 $b\bar{b}$ & 20.1 mb  & 10.75 mb & 18 (15.5) mb & 8.9 mb\\
\hline
\hline
\end{tabular}}
%\end{center}
\label{tabhq}
\end{table}

Research in the field of QCD at high parton density deals both
with fundamental theoretical issues, such as unitarity of strong
interactions at high energies, and with the challenge of
describing experimental data coming, at present, from HERA and
RHIC and expected exciting physics of forthcoming experiments at
LHC.
Over the past few years much theoretical effort has been devoted
towards the understanding of the growth of the total scattering
cross sections with energy. These studies are mainly  motivated by
the violation of the unitarity  (or Froissart) bound by the
solutions of the linear perturbative DGLAP  and BFKL  evolution
equations. Since these evolution equations predict that the cross
section rises obeying a power law of the energy, violating the
Froissart bound, unitarity corrections are expected to stop its
further growth.

In this paper we have presented a brief review
of the basic concepts present in the high density approaches and
discussed some aspects of the rapidly developing field of QCD at
high parton density in $ep$, $eA$ and $AA$ collisions.
 The successful description of all
inclusive and diffractive deep inelastic data at the collider
HERA, as well as some recent results from RHIC,  by saturation
models suggests that these effects might become important in the
energy regime probed by current colliders. In particular, the
remarkable property of geometric scaling verified in the data
indicate that the experiments are in a kinematical region which
probe QCD in the non-linear regime of high parton density.
 These results show that the transition
between the linear and non-linear regimes in $eA$ processes at high
energies will occur in a perturbative regime, justifying
perturbative QCD approaches. Our recent studies shown that an
alternative for  $eA$ colliders is the study of saturation effects
in ultraperipheral heavy ion collisions.

\section*{Acknowledgments}
  This work was partially
financed by the Brazilian funding agencies CNPq and FAPERGS.

\end{document}